\documentclass[usenatbib]{mn2e}

\usepackage{epsfig}
\usepackage{graphicx}
\usepackage{tabularx}
\usepackage{amsmath}
\usepackage{amssymb}
\usepackage{amstext}
\usepackage{amsfonts}
\usepackage{xspace}

\newcommand{\kms}[0]{\unskip\ensuremath{\,\textup{km\,s}^{-1}}}
\newcommand{\cm}[1]{\unskip\ensuremath{\,{\rm cm}^{#1}}}
\newcommand{\s}[1]{\unskip\ensuremath{\,{\rm s}^{#1}}}
\newcommand{\bm}[1]{\ensuremath\mathbf{#1}}
\newcommand{\dd}{\ensuremath\text{d}}
\newcommand{\eg}[0]{{\em e.g.}\xspace}
\newcommand{\etc}[0]{{\em etc.}\xspace}
\newcommand{\ie}[0]{{\em i.e.}\xspace}
\newcommand{\Pa}[0]{{\sc PI}\xspace}

\title[Dynamics of buoyant bubbles...]{Dynamics of buoyant bubbles in clusters of galaxies}

\author[Georgi Pavlovski et al.]%
{Georgi Pavlovski$^1$%
\thanks{Email: gbp@phys.soton.ac.uk (GP)},
Christian R. Kaiser$^1$,
Edward C.D. Pope$^{1,2,3}$\\
$^1$ School of Physics and Astronomy, University of Southampton,
Southampton SO17 1BJ, U.K.\\
$^2$ School of Engineering Sciences, University of Southampton,
Southampton SO17 1BJ, U.K.\\
$^3$ School of Physics and Astronomy, University of Leeds, Leeds, 
LS2 9JT, U.K.
}
\date{Accepted .... . Received .... ; in original form .... }

\begin{document}

\maketitle

\begin{abstract}
We present a phenomenological model of the dynamics of buoyant
bubbles in the atmosphere of a cluster of galaxies.  The derived
equations describe velocity, size, mass, temperature and density of
the buoyant bubbles as functions of time based on several simple
approximations.  The constructed model is then used to interpret
results of a numerical experiment of heating of the cluster core
with buoyant bubbles in a hydrodynamical approximation (\ie in the
absence of magnetic fields, viscosity, and thermal diffusion).  Based
on the model parameters we discuss possible limitations of the
numerical treatment of the problem, and highlight the main physical
processes that govern the dynamics of bubbles in the intracluster
medium.
\end{abstract}

%-------------------------------------------------------------------------


%=========================================================================
\section{Introduction}
\label{int}
%-------------------------------------------------------------------------
Depressions of the surface brightness in X-ray images of clusters of
galaxies have been identified as low density and high temperature
plasma bubbles \cite[see, \eg,][]{McNamara00,McNamara01,Fabian06},
created by outflows from the central active galactic nucleus (AGN).
Due to their high heat content these bubbles are thought to be one of
the main heating sources of the cluster core, and the key ingredient
in the solution of the cooling flow problem \citep{Fabian94}.

The physical process of the deposition of the heat from the bubbles
into the ambient ICM is not well understood.  It is partly because we
do not know all the relevant physical properties of the ICM (its
turbulent velocities, viscosity, \etc, see \cite{Schekochihin05} for
further discussion), and partly because theoretical models based on
the numerical simulations with all the relevant physics are very
complex, and more sophisticated numerical models, \eg including
realistic magnetic fields \citep{Ruszkowski07}, are not yet very
common.

In order to get a better understanding of the physical processes that
determine the behaviour of AGN-blown bubbles in the ICM we start from
a basic hydrodynamical model.  In the present article we analyse the
physics behind the results of numerical simulations of the evolution
of buoyant bubbles in the atmosphere of a cluster of galaxies in the
absence of magnetic fields, viscosity, and thermal diffusion.  We
derive equations that describe velocity, size, mass, temperature and
density of the buoyant bubbles as functions of time based on several
simple approximations.  This approach highlights the important
hydrodynamical effects, and helps to understand and discuss
limitations of the numerical framework for the description of AGN
bubbles.  It should be viewed as a first step in the modelling
process, that can isolate the phenomena caused by the simplest
hydrodynamical effects.

The structure of the article is as follows: in section~\ref{num} we
outline the setup of the numerical simulation, which we later use to
fit our model parameters; in section~\ref{phn} we describe the
vortex-ring model for AGN bubbles, derive the equations that determine
the evolution of the parameters of the bubbles, and determine values
for the free phenomenological parameters of the model from comparison
with the numerical data; in section~\ref{ana} we analyse the physical
properties of the model, and how numerical artifacts can affect them;
in section~\ref{con} we summarise the main findings and conclusions.
Appendices~\ref{s:sysodes}, \ref{s:AddedMass}, and
\ref{s:velVortexRing} provide some background information about the
hydrodynamics of the vortex rings.

%=========================================================================
\section{Numerical Experiment}
\label{num}
%-------------------------------------------------------------------------
In our previous work \cite[][ hereafter {\sc PI}]{Pavlovski06a} we have
demonstrated that the evolution of buoyant bubbles in the ICM is
influenced by the hydrodynamical Kutta-Zhukovsky forces, which exist
due to the circulation of the plasma in the space occupied by the
bubble.  One of the goals of the present work is to provide a
quantitative description of the role of these forces in the overall
dynamics of the bubbles.

\begin{figure}
\centering\includegraphics[width=0.8\linewidth]{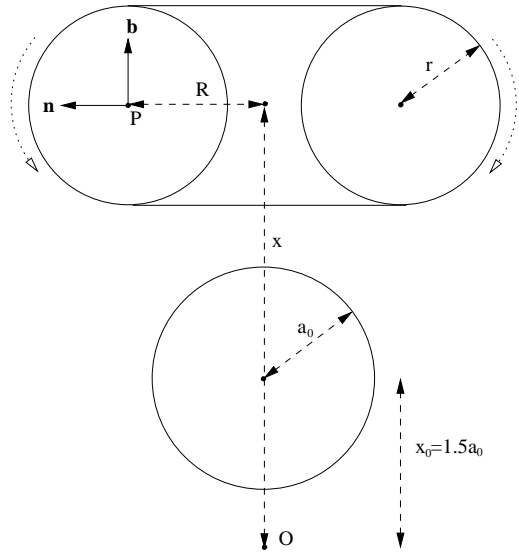}
\caption[Two morphologies of a bubble]{Initial size and position of
  the spherical bubble, and geometry and size of the torus-shaped
  bubble during the second stage of the ascent (not to scale).  Point
  $O$ marks the centre of the cluster, $a_0$ is the initial size of
  the bubble, $x_0=1.5a_0$ is its initial position.  $R$ is the radius
  of the torus, $r$ is the radius of the torus cross section, $P$ is a
  point on the central circle axis line of the torus, vectors $\bm{b}$
  and $\bm{n}$ are accordingly the bi-normal and the normal vectors of
  the central circle axis line at point $P$ (see
  Appendix~\ref{s:velVortexRing}).}
\label{f:sch}
\end{figure}

Findings in \Pa were based on the analysis of numerical experiments of
the buoyant ascent of bubbles in the ICM.  The cluster atmosphere was
modelled using the observational data for the Virgo cluster
\citep{Ghizzardi04}.  The bubbles were introduced {\em ad hoc} into an
established cooling flow, at the point when the central temperature
dropped below 1.5 keV.  The initial temperature of the bubbles was
fixed at $T=5\times10^{10}$ K (a factor of $10^3$ larger than the
temperature of the ambient ICM).  The density of the plasma inside the
bubbles was calculated from the condition of the overall pressure
equilibrium with the ambient medium (we tested two different pressure
profiles).  The initial bubbles' centres were placed symmetrically at
the distance of $1.5a_0$ from the centre of the cluster, where
$a_0=10^{22}\cm{}\approx3.24$~kpc was fixed as their initial radii,
see Fig.~\ref{f:sch}.  The size of the computational domain with
periodic boundary conditions was $10^{24}\cm{}$.  The computation was
performed using the {\sc FLASH} AMR hydrocode with 8 levels of
refinement, resulting in a smallest cell size of
$\approx4.88\times10^{20}\cm{}$.

Our motivation for the choice of $a_0$ was the following.  If $E$ is
the amount of energy released by an AGN, and $E_0 = f E$ is the
fraction of the energy used to heat the ICM during the inflation of
the bubble, then the volume of such a bubble is,
\begin{equation}
\label{e:HS1}
V_0 = E_0 (\gamma - 1) / P,
\end{equation}
where $P$ is the ICM pressure, and $\gamma=5/3$ is the ratio of the
specific heats.  For typical AGN values this gives the size of the
bubble as,
\begin{equation}
\label{e:HS2}
\begin{split}
a_0 &= 3.25 [\text{kpc}]\biggl[ \left( \frac{ f }{ 0.025 } \right)
\left( \frac{ L }{ 10^{42} [\text{erg}\,\text{s}^{-1}] } \right)\times
\\ &\left( \frac{ \tau }{ 10^8 [\text{yr}] } \right) \left( \frac{
10^{-3} [\text{cm}^{-3}] }{ n_\text{amb} } \right) \left( \frac{ 10^8
[\text{K}] }{ T_\text{amb} } \right) \biggr]^{1/3}
\end{split}
\end{equation}
where $L$ is an average AGN bolometric luminosity, and $\tau$ is an
average time between bubble injections, $n_\text{amb}$ is the number
density of the ICM, $T_\text{amb}$ its temperature, and $f$ sets the
efficiency of the thermal coupling of the outflow with the ICM.  We
here assume $f$ to lie in the range of a few percent
\citep{Sijacki06}.

%=========================================================================
\section{Phenomenological Model}
\label{phn}
%-------------------------------------------------------------------------
Analysis of the dynamics of large scale thermals and plumes in the
atmosphere of the Earth \cite[both natural and created during
(nuclear) test explosions, see, \eg,][]{Turner57, Morton60, Woods97,
Zhou01} suggests that the evolution of hot bubbles in a stratified
atmosphere can be split into two stages.  \cite{Hristianovich54}
proposed that during the first, relatively short stage of the ascent
of a hot bubble it can be treated as remaining roughly spherical,
while during the second stage its shape is best approximated by a
torus. During the second stage the dynamics of the bubble is
influenced by the velocity field around the bubble, which corresponds
to that of a vortex ring.  It is due to this velocity field that the
morphology of the bubble changes (see \Pa for more details).
Qualitative analysis of numerical simulatio of the hot bubbles in the
ICM in the absence of magnetic fields done by other research groups
\citep{Gardini06, Reynolds05, Brueggen02} also support this picture.

The morphological similarity of these (vastly different in size)
bubbles is hardly surprising as the physical processes that govern
them are very similar in the framework of a purely hydrodynamical
model of the ICM and AGN bubbles.  Such model is of course an
oversimplification of the real phenomenon.  AGN bubbles are filled
with a relativistic magnetised plasma, which has an effect on the heat
capacity of the plasma, and affects the diffusion properties at the
bubble/ICM interface \citep{Ruszkowski07,Jones07}.  Here, however, we
will assume that the magnetic fields are dynamically not important,
and will view the plasma inside bubbles as thermal, simply having a
much larger temperature.

In this rough approximation, both jet-blown ICM bubbles and buoyant
thermals in the Earth's atmosphere are characterised by high
temperature and density contrast, being, on average, in overall
pressure equilibrium with the ambient medium.  Any departure from
pressure equilibrium is smoothed out on the time scale of the order of
$\sim a/c$, where $a$ is the size of the bubble, and $c = \sqrt{\gamma
P/\rho}$ is the adiabatic speed of sound of the gas inside the bubble.
The dynamical time scale, on the other hand, is $\sim a /
c_\text{amb}$, where $c_\text{amb}$ is the adiabatic speed of sound of
the ambient medium.  In the case of hot, low density bubbles,
$c_\text{amb} /c \ll 1$, and we will assume that pressure equilibrium
is maintained at all times.  Finally, both hot bubbles in the ICM and
hot atmospheric bubbles are rising due to the presence of the external
gravitational field.

Based on this qualitative comparison we present a phenomenological
model of the evolution of the buoyant bubbles in the ICM based on the
work of \cite{Onufriev67} \citep[for a recent review see][in
Russian]{Bel00}.

%------------------------------------------------------------------
\subsection{Growth of the spherical bubble}
\label{ss:GrowthOfTheBubble}
%------------------------------------------------------------------

In this section we will estimate the change of the size of a spherical
bubble as it ascents in a stratified atmosphere of a cluster.  In a
static atmosphere the only physical processes that affect the volume
of the bubble are the entrainment of the colder ambient plasma (at the
same pressure), change of the ambient pressure (as the bubble ascents
the ambient pressure falls), and energy loss due to bremsstrahlung
radiation.  We can estimate the change of the volume due to these
processes from the following simple argument.

Consider a hot plasma bubble ascending buoyantly a distance $\dd x$
during the time period d$t$, while a mass d$M$ of the ambient gas at the
temperature $T_\text{amb}$ is entrained into it.  Note, that the
entrained plasma needs to be accelerated to the velocity $\dd x/ \dd t
= \dot{x}$ to travel with the bubble.  However, the fraction of the
total energy of the bubble which goes into the increase of the kinetic
energy of the entrained plasma is going to be much smaller than the
fraction of energy which goes into heating of it to the new higher
temperature $T$, provided that the ascent speed of the bubble is
smaller than the speed of sound in the ambient ICM, $|\dot{x}| <
c_\text{amb}$.  Here we will ignore the kinetic energy of the
entrained mass.

It is obvious that,
\begin{equation}
\label{e:GOTB1}
\frac{\dd V}{V} = \frac{\dd M}{M} - \frac{\dd \rho}{\rho}.
\end{equation}
To estimate the right hand side of (\ref{e:GOTB1}) we note that in
pressure equilibrium $\dd \rho/\rho=-\dd T/T$, and the temperature of
the mixture, $T$, could be found from the equation of internal
energy conservation,
\begin{equation}
\label{e:GOTB2}
T_\text{amb} \dd M + T M = (T + \dd T) (M + \dd M).
\end{equation}
where we have ignored differences in the heat capacities of the
ambient plasma and plasma inside the bubble. Hence,
\begin{equation}
\label{e:GOTB3}
\begin{split}
(\zeta - 1)\frac{\dd M}{M} &=\frac{\dd T}{T} \\ &=-\frac{\dd
\rho_1}{\rho}, \\
\end{split}
\end{equation}
where $\dd \rho_1$ is the change of density of the bubble due to the
entrainment of the mass $\dd M$ of the ambient plasma at temperature
$T_\text{amb}$, and $\zeta$ is the {\em contrast parameter},
\begin{equation}
\label{e:GOTB4}
\zeta = \frac{ \rho }{ \rho_\text{amb} } = \frac{ T_\text{amb} }{ T },
\end{equation}
which can be related to the X-ray detectability of the bubbles;  \eg, 
the ratio of the X-ray fluxes due to bremsstrahlung, $\propto
\rho^2 T^{1/2}$, is equal to $\zeta^{3/2}$.

Using equations (\ref{e:GOTB4}) and (\ref{e:GOTB1}) we get the
following equation for the change of volume due to the entrainment in
the adiabatic homogeneous ICM,
\begin{equation}
\label{e:GOTB5}
\begin{split}
\frac{\dd V}{V} &= \zeta \frac{\dd M}{M}\\ &= \dd \zeta + \zeta
\frac{\dd V}{V},
\end{split}
\end{equation}
which has the simple solution,
\begin{equation}
\label{e:GOTB6}
\begin{split}
\frac{V}{V_0} &= \frac{1-\zeta_0}{1-\zeta}\\ &=
\left(\frac{a}{a_0}\right)^3,
\end{split}
\end{equation}
where $V_0$ is the initial volume of the bubble,
$a_0=10^{22}\cm{}=3.24$~kpc its initial radius and $\zeta_0$ is the
initial contrast parameter.  Equation (\ref{e:GOTB6}) is due to
\cite{Hristianovich54}.

In our simulations bubbles are injected with the $T_0 \approx
10^{10}$~K, and $\zeta_0 \approx 10^{-4}$.  When the temperature of
the bubbles falls to, {\it e.g.}, $T=2\times10^7$~K, $\zeta=0.5$, the
corresponding relative change of the size of the spherical bubble (due
to the entrainment) is $a/a_0 \approx 1.26$.  It is worth noting that
we have implicitly assumed here that the entrained material is mixed
fully on time scales shorter than $|a_0/\dot{x}|$, and the bubble is
homogeneous at all times.  This assumption is generally not correct.
The boundary layer of the bubble is entraining material, which is then
getting mixed via turbulent motions or diffusion with the plasma in
the bubble's centre.  In this case the temperature of the bubble in
not homogeneous: the temperature at the boundary of the bubble is
lower then the temperature at the centre.  When determining the size
of the bubble from observations (or numerical data), depending on an
{\it ad hoc} threshold in some gas property used to find the boundary
of the bubble, it can appear that the bubble is getting smaller, rather
then larger, as ambient ICM is entrained into it.

It is straightforward to show that the change of size of the bubble
due to the change of the ambient pressure, and the loss of energy by
bremsstrahlung radiation is small -- generally not exceeding ten
per cent.

As the bubble ascents a distance $\dd x$, it reaches a layer with
lower ambient pressure and adiabatically expands,
\begin{equation}
\label{e:GOTB7}
\dd \rho_2 = \rho \frac{\dd P}{\gamma P}.
\end{equation}
Assuming pressure equilibrium and using equation (\ref{e:GOTB1}) we get,
\begin{equation}
\label{e:GOTB8}
\frac{V}{V_0} = \left(\frac{P_0}{P}\right)^{1/\gamma}.
\end{equation}
For a distance $x=3.25a_0$ travelled by the bubbles in our numerical
experiment (see PI), we get $a/a_0\approx1.12$.

The plasma inside the bubble emits bremsstrahlung radiation and cools.
The density inside the bubble changes $\dd \rho/\dd \rho = -\dd T/T=
-\dd E / E$, where $\dd E$ is the loss of internal energy due to
bremsstrahlung.  The resulting change of the volume is,
\begin{equation}
\label{e:GOTB9}
\frac{V}{V_0}=1 - \frac{\Delta E}{E},
\end{equation}
where $\Delta E$ is the total loss of internal energy due to
bremsstrahlung during the ascent of the bubble.  We can estimate the
upper limit of $\Delta E/E$ using the values of density and
temperature of the bubble near the end of its ascent. Using data from
the numerical experiment (see PI): $\Delta
t\approx7.7\times10^{14}\s{}$, $\rho\approx10^{-26}\text{~g}\cm{-3}$,
$T\approx10^8$~K, we get $0.99<a/a_0\leq1$.

All of the mechanisms described above change the size of the bubble
isotropically, and as a result the radius of the bubble towards the
end of the ascent $a\approx1.35a_0$. To account for the preferential
sideways enlargement of the bubbles, eventually leading to formation
of a torus, and reconcile this analysis with the numerical data, we
need to consider the main forces acting on the bubble during the
ascent.

%------------------------------------------------------------------
\subsection{Buoyant Vortex Ring}
\label{ss:BuoyantVortexRing}
%------------------------------------------------------------------

\begin{figure*}
\centering\includegraphics[height=6cm]{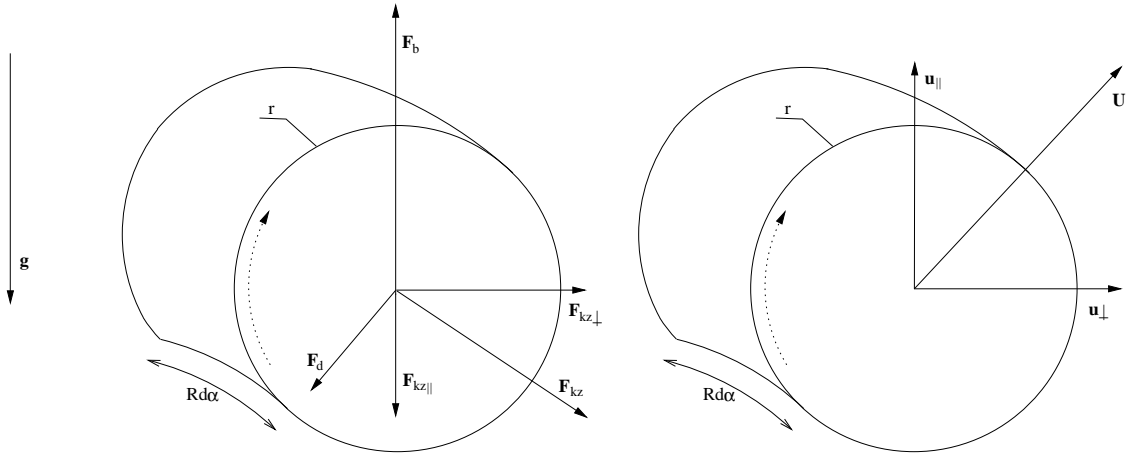}
\caption[Vortex Ring Schematic]{Right panel: forces acting on a
segment (element) of the vortex ring. $F_\text{b}$ -- buoyancy, $F_\text{d}$ --
drag, $F_\text{kz}$ -- Kutta-Zhukovsky force, $F_{\text{kz}||}$,
$F_{\text{kz}\bot}$ -- components of the Kutta-Zhukovsky force.  Left
panel: velocity components. $U$ -- total velocity vector of the centre
of mass of the vortex ring element, $u_{||}$ -- vertical component of
the velocity vector, $u_{\bot}$ -- horizontal velocity (expansion) due
to the component Kutta-Zhukovsky force $F_{\text{kz}\bot}$.}
\label{f:BVR1}
\end{figure*}

Sections \ref{ss:BuoyantVortexRing}, \ref{ss:EquationOfMotion},
\ref{ss:HeatBalance}, \ref{ss:size} are based on the work of
\cite{Onufriev67} and \cite{Hristianovich54}.

The main force that determines the ascent of the AGN blown bubbles in
the ICM is created by buoyancy due to the difference in densities of
plasma inside the bubbles and the ambient ICM.  During the ascent the
mass entrainment creates friction forces which lead to the formation
of a circulational motion of gas inside the bubble and around its
surface.  As the speed of the ascent grows, the spherical symmetry
breaks down.  The initially spherical bubble expands sideways, since the
pressure on the ``top'' and the ``bottom'' of the bubble becomes
larger than the pressure on the sides (the velocity of the flow is
larger on the sides).  This process leads to the change of the shape
of the bubble from spherical into torus-like, see Fig~\ref{f:sch}.
Inside the torus the gas rotates around the central circle axis line,
and outside the torus a circulation flow is formed.

The circulation of plasma around an element of the torus results in
a Kutta-Zhukovsky force, $F_{\text{kz}}$, see Fig.~\ref{f:BVR1}, which
is perpendicular to the overall velocity, $U$, of the element.
One of the components the Kutta-Zhukovsky force, $F_{\text{kz}\bot}$,
is perpendicular to the direction of the ascent ($\bot$-direction),
and acts to enlarge the torus radius $R$.  Another component of the
Kutta-Zhukovsky force, $F_{\text{kz}||}$, is pointing in the direction
opposite to the direction of the ascent ($||$-direction) and reduces
the velocity of the ascent.

The drag due to the viscosity and the kinetic energy of the wake
underneath the bubble results in the drag force $F_\text{d}$, which
also acts to slow down the ascent.  Note taht the $F_\text{d}$ acts in
the direction opposite $U$, and hence not directly opposite to
$F_\text{b}$.

%------------------------------------------------------------------
\subsection{Equation of motion}
\label{ss:EquationOfMotion}
%------------------------------------------------------------------

The rate of change of momentum of an element of the vortex ring plus
the associated rate of change of momentum of the surrounding medium
(the latter is modelled as the added-mass, see
Appendix~\ref{s:AddedMass}) is equal to the sum of the forces acting on
the vortex ring element, see Fig.~\ref{f:BVR1}.  We can approximate the
drag force using the formula applicable for a cylinder with a
circular cross section of radius $r$ moving through a fluid
\citep{Tritton88},
\begin{equation}
\label{e:EOM1}
F_\text{d} = \frac{1}{2}C_\text{d}\rho_\text{amb}U^2\pi r^2,
\end{equation}
where $C_\text{d}$ is a constant, $\rho_\text{amb}$ is the ambient
density, $U$ is the velocity of the vortex ring element, and $r$ is the
radius of its cross section.  The experimental value of the constant
$C_\text{d}$ varies in range $0.1\ldots100$, and depends on the
Reynolds number, $R\!e$, of the flow.  Generally it is larger for
flows with smaller $R\!e$.  It is difficult to predict its value for
our case, since the ``cylinder'' in this case is an element of the
vortex ring, which is not a solid body, as it is under-dense
compared to the ambient medium.  The main motivation for using equation
(\ref{e:EOM1}) is purely geometrical.

The Kutta-Zhukovsky force is proportional to the circulation,
$\Gamma$, (see appendix~\ref{s:velVortexRing} and \Pa) of the flow
outside the vortex ring, and is perpendicular to the velocity, $U$, of
the element of the torus (vortex ring),
\begin{equation}
\label{e:EOM2}
F_\text{kz} = \Gamma \rho_\text{amb} U R \dd\alpha,
\end{equation}
where $R$ is radius of the torus, and $\alpha$ is the central angle.

By projecting the vector quantities on the $||$ and $\bot$ directions, we
get the following equations of motion,
\begin{equation}
\label{e:EOM3}
\begin{split}
\frac{\dd}{\dd t}\left( u_{||} \left( M + M'\right)\right) &= g V
\rho_\text{amb}(1-\zeta) - 2\pi\rho_\text{amb}\Gamma R u_{\bot} \\
&-\frac{1}{2}C_\text{d}\pi r^2 u_{||}U,
\end{split}
\end{equation}
and
\begin{equation}
\label{e:EOM4}
\frac{\dd}{\dd t}\left( u_{\bot} \left( M + M'\right)\right) = -
2\pi\rho_\text{amb}\Gamma R u_{||} -\frac{1}{2}C_\text{d}\pi r^2
u_{\bot}U,
\end{equation}
where $M'=2\pi^2r^2R\rho_\text{amb}$ is the added-mass for the torus (see
Appendix~\ref{s:AddedMass}), $g$ is the gravitational acceleration,
and
\begin{equation}
\label{e:EOM5}
U = \sqrt{u_\bot^2 + u_{||}^2}
\end{equation}
is the velocity of the torus element.

%------------------------------------------------------------------
\subsection{Heat balance}
\label{ss:HeatBalance}
%------------------------------------------------------------------
Entrainment of the cooler ambient plasma changes the density of the
bubble according to equation (\ref{e:GOTB3}), while the bubble also
expands due to the fall of the ambient pressure as given by equation
(\ref{e:GOTB9}).  From the definition of the contrast parameter,
\[
\frac{\dd\zeta}{\zeta}=
\frac{\dd\rho}{\rho}-\frac{\dd \rho_\text{amb}}{\rho_\text{amb}},
\]
we get the following formula for the change of contrast due to the mass
entrainment and the change in the ambient pressure and density,
\begin{equation}
\label{e:HB1}
\frac{\dd\zeta}{\zeta} = (1-\zeta)\frac{\dd M}{M} + \frac{\dd
  P}{\gamma P} - \frac{\dd \rho_\text{amb}}{\rho_\text{amb}},
\end{equation}
which implies the following differential equation,
\begin{equation}
\label{e:HB2}
\frac{\dd\zeta}{\dd t} = \zeta\left[ (1-\zeta)\frac{1}{M}\frac{\dd
M}{\dd t} + \dot{x}\left( \frac{1}{\gamma P}\frac{\dd P}{\dd x}-
\frac{1}{\rho_\text{amb}}\frac{\dd \rho_\text{amb}}{\dd x} \right)\right].
\end{equation}
The last term in (\ref{e:HB2}), which is enclosed in round
parenthesis, is a measure of the deviation of the ambient atmosphere
from adiabatic conditions.  For an adiabatic atmosphere
$\dd\rho_\text{amb}/\rho_\text{amb}=\dd P/(P\gamma)$, and the term is
exactly zero.

%------------------------------------------------------------------
\subsection{Mass entrainment}
\label{ss:ME}
%------------------------------------------------------------------
The entrainment of the ambient ICM into the bubble happens due to
growing fluid instabilities on the surface of the bubble.  The
boundary layer of the buoyant bubble is corrugated by the
Rayleigh-Taylor (RT) instability and becomes turbulent.  The
interpenetration distance $h$ of the light fluid into the heavy fluid
in a classical RT instability scenario can be written as \citep[see,
\eg,][]{Sharp84},
\begin{equation}
\label{e:ME5}
h = C_\text{m} A\!t g t^2,
\end{equation}
where
$A\!t=(\rho_\text{amb}-\rho)/(\rho_\text{amb}+\rho)=(1-\zeta)/(1+\zeta)$
is the Atwood number, $g$ is gravitational acceleration, $t$ is time, and
$C_\text{t}$ is a constant \citep[see also][ for a discussion of
numerical methods for mass diffusion and values of the
constant]{Liu07}.  Using (\ref{e:ME5}) we get the following equation
for the mass transfer rate due to the RT instability,
\begin{equation}
\label{e:ME6}
\frac{1}{M}\frac{\dd M}{\dd t} = C_\text{m}
\frac{1-\zeta}{1+\zeta}\frac{U}{r},
\end{equation}
were we have substituted $U$ for $gt$ as it is a more appropriate
measure of the velocity of the bubble/ICM interface in our case.

%------------------------------------------------------------------
\subsection{Size of the bubble}
\label{ss:size}
%------------------------------------------------------------------
Using the formula for the mass of the torus,
\begin{equation}
\label{e:st0}
M=2\pi^2Rr^2\rho
\end{equation}
it is straightforward to write an equation for the change of the
sizes, $R$ and $r$, of the torus,
\begin{equation}
\label{e:st1}
\frac{\dd r}{\dd t} = \frac{r}{2}\left[ \frac{1}{M}\frac{\dd M}{\dd t}
-\frac{1}{R}\frac{\dd R}{\dd t} -\frac{1}{\zeta}\frac{\dd \zeta}{\dd
t} -\frac{\dot{x}}{\rho_\text{amb}}\frac{\dd \rho_\text{amb}}{\dd
x}\right],
\end{equation}
or, equivalently,
\begin{equation}
\label{e:st2}
\begin{split}
\frac{\dd R}{\dd t} &= u_\bot,\\ \frac{\dd r}{\dd t} &=
\frac{r}{2}\left[ \frac{1}{M}\frac{\dd M}{\dd t} -\frac{u_\bot}{R}
-\frac{1}{\zeta}\frac{\dd \zeta}{\dd t}
-\frac{\dot{x}}{\rho_\text{amb}}\frac{\dd \rho_\text{amb}}{\dd
x}\right].\\
\end{split}
\end{equation}

%------------------------------------------------------------------
\subsection{Circulation}
\label{ss:cirl}
%------------------------------------------------------------------
The circulation of the flow around the element of the torus is
determined by the angular velocity of the fluid inside it.  As the
mass of the torus grows, the angular velocity of the fluid inside it
decreases.  Any disconnection between the inner and the outer
circulation flows (\ie difference in their angular velocity) leads to
the production of vortices near the surface of the torus, which then are
shredded away, slowing down the rotation of the flow further.  It is
reasonable to assume that the angular velocity is directly
proportional to the circulation of the flow around the element of the
torus, $\omega\propto\Gamma$.  Consider a cylinder of radius $r$
rotating in a fluid with kinematic viscosity $\nu$ around its axis of
symmetry.  The torque due to viscous stresses \citep[][ \S 18]{LandauIV}
per unit length of this cylinder is
$8\pi^2\nu\rho_\text{amb}Rr^2\omega=4\nu M\omega/\zeta$.  The torque of
the friction forces will change the angular momentum of the cylinder,
$1/2Mr^2\omega$.  This implies that,
\begin{equation}
\label{e:cirl00}
\frac{\dd }{\dd t}\left(Mr^2\Gamma\right)\propto\nu M\Gamma/\zeta.
\end{equation}
In the case of the vortex ring the role of the viscosity (friction
forces) is played by the interpenetration of the fluids across the
boundary of the torus.  The movement of the plasma across the
boundary, in other words the rate of mass entrainment, determines
whether the motion of the fluid on one side of the boundary has an
affect on the motion of the fluid on the other side of the
boundary. Formally, from dimensional arguments we can set $\nu\propto
\zeta r^2 M^{-1}\dd M/\dd t$.  Substituting the expression for $\nu$
into equation (\ref{e:cirl00}) and assuming that the rate of change of
the radius, $r^{-1}\dd r/ \dd t$, is much smaller than the rate of
change of the mass, $ M^{-1} \dd M / \dd t$, we get the following
equation for the change of circulation,
\begin{equation}
\label{e:cirl1}
\frac{1}{\Gamma}\frac{\dd \Gamma}{\dd t}=
-C_\text{c}\frac{1}{M}\frac{\dd M}{\dd t},
\end{equation}
where $C_\text{c}$ is a constant which needs to be determined from 
comparison with the data.

%------------------------------------------------------------------
\subsection{Simultaneous ODEs}
\label{ss:ode}
%------------------------------------------------------------------
The velocity of accent $\dot{x}$ is the sum of the vertical velocity
of the element, $u_{||}$, which is determined by the buoyancy and the
hydrodynamical forces, plus the self-induced vortex ring velocity,
$v_\text{ind}$, which is the result of the curvature ($1/R$) of the
vortex ring (see Appendix~\ref{s:velVortexRing} for more information),
\begin{equation}
\label{e:ode1}
\dot{x}\equiv\frac{\dd x}{\dd t} = u_{||} + v_\text{ind},
\end{equation}
where the induction velocity $v_\text{ind}$ is proportional to the
circulation and the curvature of the vortex ring, $\propto\Gamma /R$.
In the present model we will use the expression for $v_\text{ind}$
given by equation (\ref{e:ATBB4}) in Appendix~\ref{s:velVortexRing}.

Together equations (\ref{e:ode1}), (\ref{e:cirl1}), (\ref{e:st2}),
(\ref{e:st0}), (\ref{e:ME6}), (\ref{e:HB2}), (\ref{e:EOM5}),
(\ref{e:EOM4}), (\ref{e:EOM3}) form a closed system of ordinary
differential equations (ODEs), which determines the values of eight
physical parameters, $u_\bot$, $u_{||}$, $R$, $r$, $x$, $\Gamma$, $M$,
and $\zeta$ as functions of time. Three phenomenological parameters,
$C_\text{d}$, $C_\text{m}$, and $C_\text{c}$, need to be determined
by comparison with the numerical data.

%------------------------------------------------------------------
\subsection{Solution}
\label{ss:sol}
%------------------------------------------------------------------
The simultaneous ODEs in their dimensionless form are written in full
in Appendix~\ref{s:sysodes}.  To determine the values of the
phenomenological constants, $C_\text{c,d,m}$, we have fitted the
solution of the ODEs to the data from our numerical experiment {\sc C}
described in \Pa.  As the initial condition we have selected a point
when the bubble shape changes from spherical to toroidal.  In the
simulation this happens approximately 16~Myr after the injection of the
bubble, and $t_0=722$~Myr after the start of the simulation.  For 
reference, the dimensionless time unit is approximately $8.1$~Myr, the
velocity unit is $3.9\times10^7\kms$, the distance unit is
$10^{22}\cm{}$, and the mass unit is $9.8\times10^{40}$~g. The initial
values of the physical parameters at $t_0$ were measured using the
averaging techniques described in \Pa.  Their numerical values were
found to be as follows, $u_\bot(t_0)=0.26$, $u_{||}(t_0)=0.42$,
$R(t_0)=0.99$, $r(t_0)=0.72$, $x(t_0)=2.56$, $\Gamma(t_0)=2.71$,
$M(t_0)=2.4\times10^{-2}$, $\zeta(t_0)=2.4\times10^{-3}$.

We solve the ODEs numerically using the adaptive step Bulirsch-Stoer
method \citep{PTVF02}.  To find values of the phenomenological
constants we minimised the function,
\begin{equation}
\label{e:sol0}
\sum_{j=1}^{42}\sum_{i=1}^{5}\frac{(X_i(t_j) - Y_i(t_j))^2}{(\Delta X_i(t_j))^2},
\end{equation}
where $t_j$ are times of the snapshots of the computational grid
produced during the numerical simulation (we have 42 snapshots in
total), $X_i=\{M,R,r,x,\zeta\}$ are the parameters of the bubble
(torus) calculated from these snapshots, $\Delta X_i$ are their
estimated errors, and $Y_i(t_j)$ are the parameters of the bubble
(torus) from the solution of the simultaneous ODEs for the times
$t_j$.  The estimated errors, $\Delta X_i$, were calculated from one
sigma variations of the statistically determined quantities (see \Pa),
and errors due to the finite grid resolution.  We did not include into
$X_i$ parameters that rely on the estimation of velocities, since the
comparatively large and varying time between the snapshots,
$\tau(j)=t_j-t_{j-1}\neq\text{const}$, makes it difficult to determine
the values of the velocities accurately.

The best fit values for the constants were found to be
$C_\text{c}=0.01$, $C_\text{m}=4$, $C_\text{d}=24$.  Due to the
uncertainties in the values of the parameters at $t_0$ (especially
velocities and circulation) the resulting values of the constants are
not determined precisely.  The values of the constants are going to be
different if different initial conditions are used.  By varying the
initial parameters by ten percent, we found that the corresponding
variation of the values of the constants was approximately limited to
the ranges, $C_\text{c}\pm0.05$, $C_\text{m}\pm1$,
$C_\text{d}\pm10$.

%=========================================================================
\section{ANALYSIS OF SOLUTION}
\label{ana}
\begin{figure}
\centering\includegraphics[width=\linewidth]{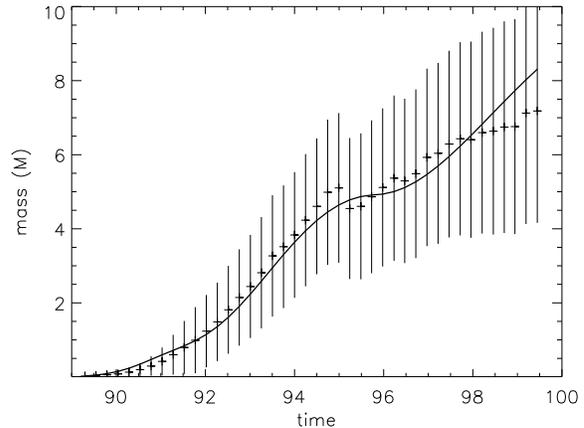}
\caption[Mass change]{Growth of the bubble mass: solid line is the best
  fit model, points with vertical error bars are measurements from
  the numerical simulation.}
\label{f:mass}
\end{figure}

\begin{figure}
\centering\includegraphics[width=\linewidth]{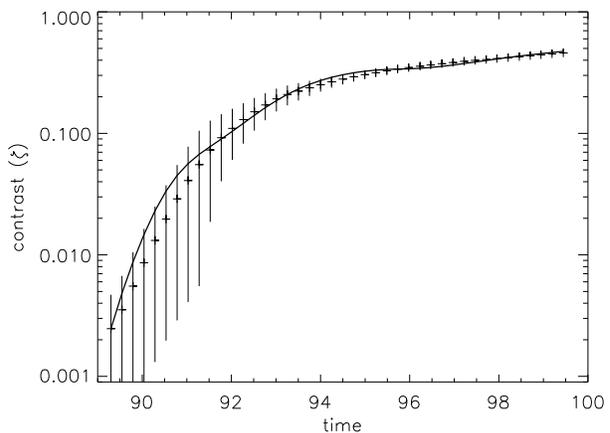}
\caption[Contrast change]{Change of the contrast parameter with time: solid
  line is the best fit model, points with error bars are measurements
  of the contrast paranter from the numerical simulation.}
\label{f:contrast}
\end{figure}

\begin{figure}
\centering\includegraphics[width=\linewidth]{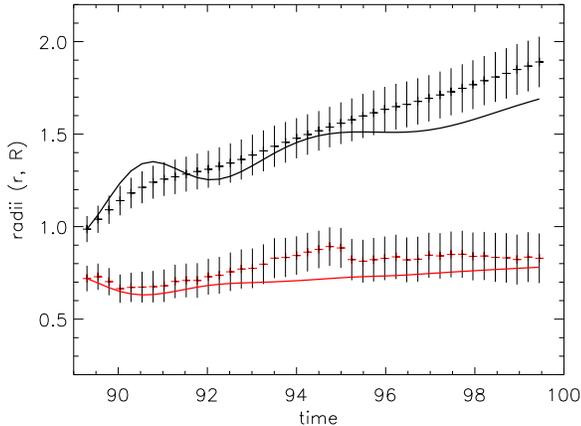}
\caption[Radii change]{Growth of the bubble: solid lines are the best
  fit models (upper curve for $R$, lower curve (red) for $r$), dashes with
  the vertical error bars are measurement from the numerical simulation.}
\label{f:radii}
\end{figure}

\begin{figure}
\centering\includegraphics[width=\linewidth]{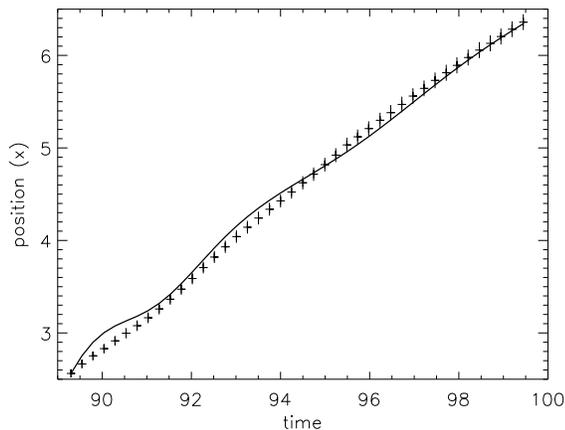}
\caption[Position]{Ascent of the bubble: solid line -- the best
  fit model, dashes with error bars -- measurement from the numerical
  simulation.}
\label{f:position}
\end{figure}

%=========================================================================
In Figs.~\ref{f:mass}, \ref{f:contrast}, \ref{f:radii}, and
\ref{f:position} we compare the evolution of the parameters determined
from the numerical simulation, $X_i$, with the prediction of our
model, $Y_i$.  It is clear that the model with the best fit constants
determined above, reproduces the data from the numerical experiment
quite well.  We achieve a good fit for the contrast parameter and the
mass of the bubble over three orders of magnitude in scale.  The fit
for the bubble radii seems to be a bit more problematic.  It is
important to note, however, that by changing the initial conditions
for the solution of the ODEs we
can get rather better fits for radii (at least by eye).  This
sensitivity to the initial conditions is not surprising.  We find that
all $Y_i$, but especially radii and position,
depend on the value of the initial circulation, $\Gamma(t_0)$. Since
the constant $C_\text{c}\approx0$, the circulation remains roughly
constant in time (conserved), and the precision of the measurement of
the initial value is reflected in the overall quality of the fit of the
parameters that depend on it directly, \ie $R$, $r$, and $x$; the
latter through the self-induced velocity of the vortex,
$v_\text{ind}$.

\begin{figure*}
\centering\includegraphics[width=0.8\linewidth]{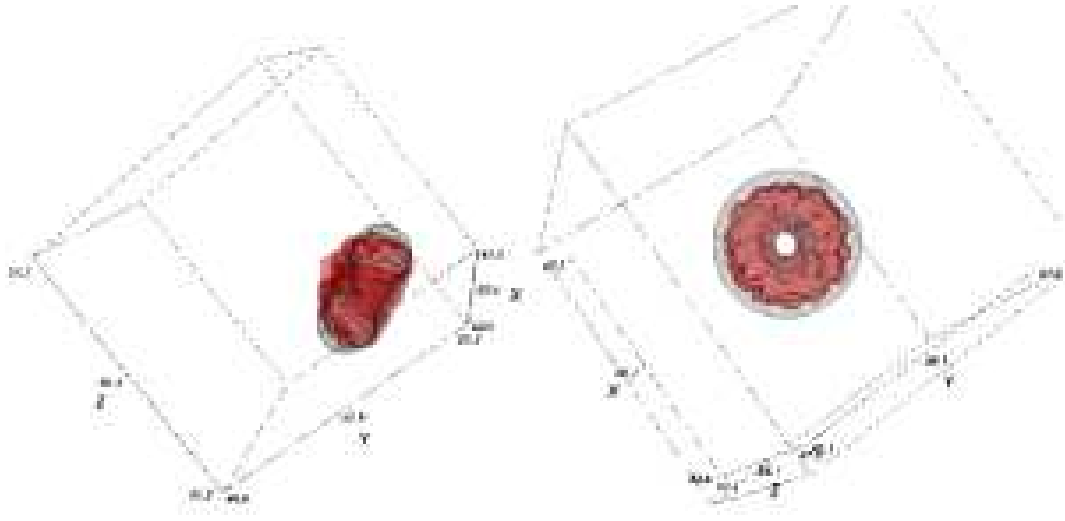}
%\centering\includegraphics[width=0.8\linewidth]{362tori.eps}
\caption[3d fit1]{The initial conditions ($t=89.1$): the
  distribution of the bubble material and the approximating torus.}
\label{f:3d1}
\end{figure*}

\begin{figure*}
\centering\includegraphics[width=0.8\linewidth]{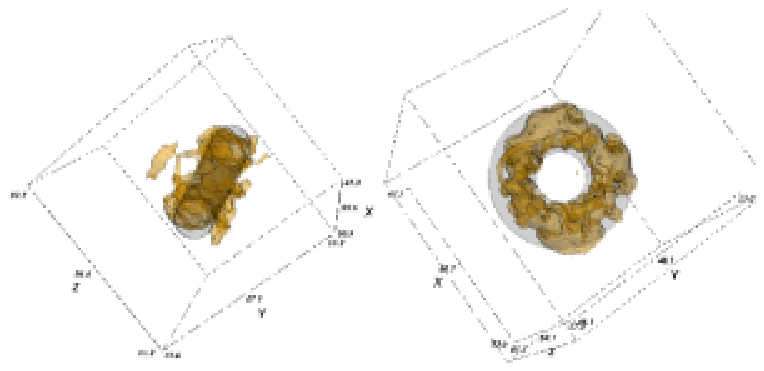}
%\centering\includegraphics[width=0.8\linewidth]{382tori.eps}
\caption[3d fit2]{Comparison of the distribution of the bubble
  material and the approximating torus at $t=93.9$.}
\label{f:3d2}
\end{figure*}

\begin{figure*}
\centering\includegraphics[width=0.8\linewidth]{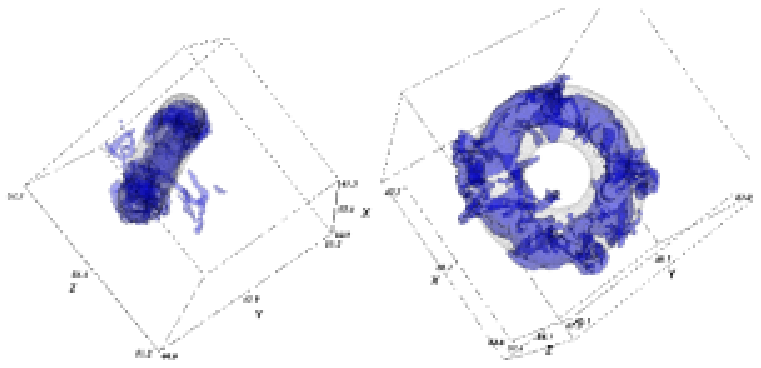}
%\centering\includegraphics[width=0.8\linewidth]{403tori.eps}
\caption[3d fit2]{Comparison of the distribution of the bubble
  material and the approximating torus at $t=99.4$.}
\label{f:3d3}
\end{figure*}

To illustrate the matching of the model to the three-dimensional
distribution of the bubble material at different times as computed in
our simulation we compare them directly in Figs.~\ref{f:3d1},
\ref{f:3d2}, and \ref{f:3d3}.  Note that the approximation of the
bubble shape with a torus at $t_0$, Fig.~\ref{f:3d1}, is not perfect.
The distribution of the bubble material, although resembling a torus,
does not have a perfectly circular cross section.  The cross section
of the bubble's torus is elliptical, and the size of the opening at
the top of the bubble is smaller than the size of the opening at the
bottom of the bubble.  At later times the matching of the shape of the
bubble to a torus with a circular cross section is much better.
Despite that, our model was able to capture the overall size and shape
of the bubbles quite well.

The solution of the ODEs suggests that the vertical velocity of the
bubbles due to the hydrodynamical forces, $u_{||}$, quickly falls due
to significant drag, as follows from the high value of
$C_\text{d}\sim30$.  In fact, we find that the ascent velocity of the
bubble, $\dot{x}$, in the current model is dominated by the
self-induced velocity of the vortex ring, $v_\text{ind}$, which is an
order of magnitude larger than the velocity $u_{||}$.  A large drag
coefficient is generally a characteristic of low Reynolds number (high
viscosity) flows.  Although we did not model any physical viscosity in
our numerical simulation, the residual numerical viscosity is always
going to be present.  It is difficult to quantify its contribution.
Qualitatively, however, we can state that the velocity of the ascent
is more significantly affected by the the self-induced velocity of the
vortex in the medium with high viscosity, and therefore large drag.
In the framework of this model the bubbles will never reach ``a
ceiling'' \ie a maximum height.  They will deposit their thermal
content along their ascent path, but even after they reach $\zeta=1$
and disappear the residual vortex will keep moving towards the
outskirts of the cluster\footnote{The drag coefficient in problem of
the rise of the cloud from a nuclear explosion is much smaller,
$C_\text{d}\sim0.5$.  \cite{Onufriev67} found that in this case the
cloud reaches the maximum height, which depends on the stratification
of the atmosphere and its initial size.}.  If the ICM has very low
viscosity, the Reynolds number is likely to be much larger than a
value of few hundred, which is characteristic for our present
simulation.  In this case the coefficient $C_\text{d}$ can be much
smaller, and $\dot{x}$ will no longer be determined by the
self-induced velocity, but rather by the balance between the buoyancy
and the vertical component of the Kutta-Zhukovsky force.

Another possible pitfall for our model is the overestimation of the
mass diffusion across the surface of the bubble from the fit to the
simulation data.  It is well known \citep{Liu07} that generic hydro
schemes tend to be super-diffusive, and yield too large values of 
coefficients like $C_\text{m}$.  We can roughly estimate the impact of
a possible reduction of the coefficient in the following way.  By
substituting equation (\ref{e:ME6}) into equation (\ref{e:HB2}), and
neglecting the properties of the atmosphere we get,
\begin{equation}
\label{e:AN0}
\frac{\dd \zeta}{\dd t}\frac{1+\zeta}{\zeta(1-\zeta)^2} = C_\text{m}\frac{U}{r},
\end{equation}
which is easily integrable if we assume that $U$ and $r$ are
constants,
\begin{equation}
\label{e:AN1}
\frac{2(\zeta-\zeta_0)}{(1-\zeta)(1-\zeta_0)}+
\log\frac{\zeta(1-\zeta_0)}{\zeta_0(1-\zeta)}=
C_\text{m}\frac{U}{r}(t-t_0).
\end{equation}
The left hand side (l.h.s.) of equation (\ref{e:AN1}) is a finite number for
any given values of the contrast parameter $0<\zeta_0<\zeta<1$
(l.h.s. $\approx27$ in our case: $\zeta=0.9$, $\zeta_0=0.001$).
Therefore, if the value of $C_\text{m}$ is reduced by a certain
factor, $f$, it will take $f$ times longer for bubbles to reach the
same contrast $\zeta$.  Given that the typical
mass diffusion constant in simulations of RT instability is
overestimated by a factor of two \citep{Liu07}, it is quite likely
that the bubbles in the ICM survive at least twice as long as a
typical simulation would tend to suggest.

The small value of the constant $C_\text{c}\approx0$ implies that the
circulation in our model is approximately conserved,
$\Gamma\approx\text{const}$, and the fluid is nearly ideal and
incompressible.  The opposite extreme $C_\text{c}=1$ corresponds to
the conservation of angular momentum, $\Gamma M=\text{const}$, see
(\ref{e:cirl1}), and would describe a bubble unaffected by the
surrounding medium.

%=========================================================================
\section{Discussion and Summary}
\label{con}
%=========================================================================
The developed model is by no means a full description of the physics of
buoyant bubbles in real clusters of galaxies.  In this work we
have tried to investigate the simplest possible approximation,
and understand how different physical properties of the bubbles are
interconnected.  Hydrodynamical approximations have been widely used to
approximate processes of heating in clusters with AGN bubbles
\citep[\eg,][]{Brueggen02,Gardini06}, and it is important to
understand the properties of such solutions in order to know their
limitations.  Magnetic fields are clearly important ingredients in
AGN physics.  They regulate deposition of heat from the bubble
into the ambient medium \citep{Jones07,Ruszkowski07} by suppressing
thermal conduction around the surface of the bubble, which
otherwise can lead to a rapid evaporation of the bubble (see \Pa).

To summarise, in this study we have developed a phenomenological
model, which describes the dynamics of buoyant AGN bubbles in the
atmospheres of clusters of galaxies.  The value of the free
phenomenological constants was determined from a comparison with the
simulation data.  The resulting fit of the prediction of the model to
the simulation data is good.  We have analysed possible outcomes that
correspond to different values of the constants.  The main points are
as follows: 1. In low Reynolds number flows the drag coefficient is
large, and the overall dynamics is determined by the circulation of
the flow around the bubble. 2. Since mass entrainment is likely to be
overestimated in numerical simulations, the bubbles can
potentially survive in the ICM much longer than predicted by a typical
simulation. 3. The conservation of circulation in our model is
a consequence of the near incompressibility of the ICM (given the
characteristic velocities), and absence of viscosity.  Note, that the
numerical viscosity alone may significantly decrease the flows
Reynolds number.

The model presented here provides a framework for the interpretation
of numerical and observational results.  This purely hydrodynamical
basis can now be extended to include more of the important physics,
\eg the effects of magnetic fields.

%=========================================================================
\section{Acknowledgements} 
This research has made use of NASA's Astrophysics Data System
Bibliographic Services.  The software used in this work was in part
developed by the DOE-supported ASC / Alliance Center for Astrophysical
Thermonuclear Flashes at the University of Chicago.

%=========================================================================
\appendix
%=========================================================================
\section{System of ODEs}
\label{s:sysodes}
%=========================================================================
Equations (\ref{e:ode1}), (\ref{e:cirl1}), (\ref{e:st2}),
(\ref{e:st0}), (\ref{e:ME6}), (\ref{e:HB2}), (\ref{e:EOM5}),
(\ref{e:EOM4}), and (\ref{e:EOM3}), which were derived in sections
\S~\ref{ss:BuoyantVortexRing} -- \ref{ss:ode} form a system of
ordinary differential equations (simultaneous ODEs).  The solution of
these equations determines values of the bubble sideways enlargement
velocity $u_\bot$, vertical velocity due to hydrodynamical forces
$u_{||}$, radial size $R$, vertical size $r$, position $x$,
circulation $\Gamma$, mass $M$, and contrast parameter $\zeta$ as
functions of time. The identified physical processes were parametrised
with the phenomenological parameters for drag $C_\text{d}$, mass
entrainment $C_\text{m}$, and circulation $C_\text{c}$.

It is convenient to work with dimensionless variables,  We since
measure sizes and distances in terms of the initial size of the bubble
$a_0$.  We can use factors $\sqrt{a_0g_0}$ for velocities and
$\sqrt{a_0/g_0}$ for time to make these variables dimensionless, where
$g_0$ is the gravitational acceleration at the starting point
$g_0=g(x_0)$.  By substituting,
\begin{equation}
\begin{split}
\{r,R,x\}\rightarrow\frac{\{r,R,x\}}{a_0},&\qquad
\{\dot{x},u_\bot,u_{||}\}\rightarrow\frac{\{\dot{x},u_\bot,u_{||}\}}{\sqrt{a_0g_0}},\\
t\rightarrow\frac{t}{\sqrt{a_0/g_0}},&\qquad
\Gamma\rightarrow\frac{\Gamma}{a_0\sqrt{a_0g_0}},\\
\end{split}
\end{equation}
into the original equations, we get, after some simple algebra, the
following ODEs describing the evolution of the parameters of the bubble:
its contrast,
\begin{equation}
\label{e:sysodes1}
\frac{\dd\zeta}{\dd t} = \zeta\left[ (1-\zeta)\frac{1}{M}\frac{\dd
M}{\dd t} + \dot{x}\frac{\dd}{\dd
x}\log(P^{1/\gamma}\rho_\text{amb}^{-1}) \right],
\end{equation}
its sizes,
\begin{equation}
\label{e:sysodes2}
\begin{split}
\frac{\dd r}{\dd t} &= \frac{r}{2}\left[ \frac{1}{M}\frac{\dd M}{\dd
t} -\frac{u_\bot}{R} -\frac{1}{\zeta}\frac{\dd \zeta}{\dd t}
-\dot{x}\frac{\dd}{\dd x}\log\rho_\text{amb}\right],\\
\frac{\dd R}{\dd t} &= u_\bot,
\end{split}
\end{equation}
its characteristic velocities,
\begin{equation}
\label{e:sysodes3}
\begin{split}
\frac{\dd u_{||}}{\dd t} &= \frac{1-\zeta}{1+\zeta} g - \frac{\Gamma
  u_\bot}{\pi(1+\zeta)r^2}\\ &+ u_{||} \left[
  \frac{1}{\zeta(1+\zeta)}\frac{\dd \zeta}{\dd t}
  -\frac{1}{M}\frac{\dd M}{\dd t}
  -\frac{C_\text{d}}{4\pi}\frac{U}{R(1+\zeta)} \right],\\
\frac{\dd u_{\bot}}{\dd t} &= \frac{\Gamma u_{||}}{\pi(1+\zeta)r^2}\\
&+ u_{\bot} \left[ \frac{1}{\zeta(1+\zeta)}\frac{\dd \zeta}{\dd t}
-\frac{1}{M}\frac{\dd M}{\dd t}
-\frac{C_\text{d}}{4\pi}\frac{U}{R(1+\zeta)} \right],\\ U &=
\sqrt{u_\bot^2 + u_{||}^2},\\
\end{split}
\end{equation}
its position (height),
\begin{equation}
\label{e:sysodes4}
\begin{split}
\dot{x} &= u_{||} + v_\text{ind},\\
v_\text{ind}&=\frac{\Gamma}{4\pi R} \biggl[\log\frac{8R}{r} -
\frac{1}{4}\\
&-\frac{12\log(8R/r)-15}{32}\left(\frac{r}{R}\right)^2\biggr],\\
\end{split}
\end{equation}
circulation of the ICM around it,
\begin{equation}
\label{e:sysodes5}
\frac{\dd \Gamma}{\dd t} = -C_\text{c}\frac{\Gamma}{M}
\frac{\dd M}{\dd t},
\end{equation}
and the mass entrainment rate,
\begin{equation}
\label{e:sysodes6}
\frac{\dd M}{\dd t} =
C_\text{m}\frac{1-\zeta}{1+\zeta}\frac{UM}{r}.
\end{equation}

%=========================================================================
\section{Added-Mass}
\label{s:AddedMass}
%=========================================================================
A body moving through a continuous medium creates a velocity field in
that medium.  A spherical bubble ascending in a static atmoshpere of a
cluster of galaxies creates a velocity field in the cluster.  The
total kinetic energy of the system, which consists of the moving
bubble plus the ambient atmosphere, is the sum of the kinetic energy
of the of the medium (the kinetic energy of the induced velocity
field), plus the kinetic energy of the bubble.  It is possible to show
that such motion can be described as a motion of a spherical bubble
with an effective mass, which consists of its own mass plus an
added-mass, which is proportional to the mass of the medium displaced
by the body.  Here we will assume that the flow is laminar, potential
(\ie non-rotational), and the bubble has a constant radius.

The velocity field in an ideal incompressible fluid with zero vorticity
can be described by the equation,
\begin{equation}
\nabla\times\bm{v}=0,
\end{equation}
or equivalently,
\begin{equation}
\label{eq:poten}
\bm{v} = -\nabla \psi.
\end{equation}
Using the incompressibility ($\nabla\cdot\bm{v}=0$) condition we get,
\begin{equation}
\label{eq:lap}
\nabla^2\psi =0,
\end{equation}
which is the well-known Laplace equation for the velocity potential,
$\psi$.

Consider a uniform flow of an incompressible, non-rotational fluid past a
sphere.  By uniform flow here we mean that at large distances from the
sphere the flow has a uniform velocity $U\bm{k}$, where $\bm{k}$ is the
unit vector (along the $z$ axis) of a Cartesian coordinate system.  We can
chose the centre of the sphere as the centre for the coordinate
system.  If the radius of the sphere is $a$, then we have to solve the
Laplace equation (\ref{eq:lap}) in the region $a\le r\le \infty$.  The
boundary condition applied to the surface of the sphere is,
\begin{equation}
\label{eq:lbound1}
\left(\frac{\partial\psi}{\partial r}\right)_{r=a}=0.
\end{equation}
The boundary condition at infinity can be taken to be,
\begin{equation}
\label{eq:lbound2}
\psi = - U r \cos\theta, \quad \text{at}\quad r\rightarrow\infty,
\end{equation}
which would give the velocity (in a spherical system of coordinates),
\begin{equation}
\bm{v}(r,\theta,\phi) = U \cos\theta \bm{e}_r - U \sin\theta
\bm{e}_{\theta},\quad\text{when } r\rightarrow\infty,
\end{equation}
where we used the relation $\bm{k} = \cos\theta \bm{e}_r - \sin\theta
\bm{e}_{\theta}$.

The general solution of the Laplace equation (\ref{eq:lap}) in
spherical coordinates is,
\begin{equation}
\begin{split}
\psi =& \sum_{l=0}^{\infty}\sum_{m=0}^{l} \left(A_lr^l +
B_lr^{-l-1}\right) P_l^m(\cos\theta)\times\\
&\times\left[S^m_l\sin\phi m + C^m_l \cos\phi m\right],
\end{split}
\end{equation}
where $P_l^m$ are the associated Legendre functions \citep{MorseII}.

The boundary conditions (\ref{eq:lbound1}) and (\ref{eq:lbound2}) can
be satisfied only for the following combination,
\begin{equation}
\psi = - U \left( r + \frac{a^3}{2r^2} \right) \cos\theta.
\end{equation}
The velocity of the flow can be easily found using (\ref{eq:poten}),
\begin{equation}
\bm{v} = U \bm{k} - U \frac{a^3}{r^3}\left(\cos\theta\bm{e}_r -
\frac{1}{2}\sin\theta\bm{e}_\theta\right).
\end{equation}

With a simple Galilean transformation we can now consider the problem
of motion of the sphere with velocity $U\bm{k}$ through a fluid at
rest (at infinity).  If we take the origin of the new coordinate
system at the instantaneous position of the centre of the sphere, then
the flow pattern around the sphere is given by,
\begin{equation}
\label{eq:svel}
\bm{v}' = - U \frac{a^3}{r^3}\left(\cos\theta\bm{e}_r -
\frac{1}{2}\sin\theta\bm{e}_\theta\right).
\end{equation}

The kinetic energy of the fluid around the sphere is,
\begin{equation}
\begin{split}
K_\text{fluid} =& \frac{1}{2}\rho\int_0^{2\pi}\int_0^\pi\int_a^\infty
\bm{v}'^2r^2\sin\theta\text{d}r\text{d}\theta\text{d}\phi\\ =&
\frac{1}{3}\pi\rho a^3 U^2\\ =& \frac{1}{2}M'U^2,
\end{split}
\end{equation}
where $M'=1/2V\rho$, $V=4/3\pi a^3$.  The total kinetic energy of the
system fluid plus the sphere is equal to the sum of the kinetic energy of
the sphere and the kinetic energy of the fluid,
\begin{equation}
K = \frac{1}{2} \left(M + M'\right)U^2,
\end{equation}
and can be interpreted as the kinetic energy of the sphere with the
{\em effective mass} of $M + M'$ -- the mass of the sphere plus an
added-mass equal to half of the displaced mass of the fluid.

A similar derivation for a cylinder yields \citep[see {\it
e.g.},][]{Rai98} $M'=V\rho$, where $V$ is the volume of the cylinder.
If we consider an element of the vortex ring to be roughly
cylindrical, then for a whole torus moving in the $||$ direction, the
added-mass also should be equal to the total mass of the displaced
fluid, $M'=V\rho$, where $V=2\pi^2r^2R$ is the volume of a torus.
A rigorous mathematical treatment of the problem \citep{Miloh78} proves
this approximation to be accurate to within a few per cent,
$1\leq M'/(V\rho)<1.0625$.

Note that the motion of {\it two} identical spheres in opposite
directions along a line connecting their centres with velocities
$U=U_1=-U_2$, yields the expression for the kinetic energy of the
fluid \citep{Lamb32},
\begin{equation}
\label{e:kin}
2K = \rho V\left(1-\frac{3}{16}\frac{a^3}{x^3}
+O\left(\frac{a^6}{x^6}\right)\right)U^2,
\end{equation}
where $V=4/3\pi a^3$ is the volume of the sphere, and $x$ is half of the
distance between the spheres. Equation (\ref{e:kin}) implies that the
resulting effective mass for a sphere in the presence of an identical
sphere moving in the opposite direction is slightly lower than that of
a single sphere.  Therefore, our approximation of the added-mass of
the torus, $M'=V\rho$, which is slightly lower than the value for a
single torus as given by calculations of \cite{Miloh78} is, in fact,
appropriate since in our cluster there are two bubbles, \ie there is a
second torus, which is moving with the same speed in the opposite
direction.

%=========================================================================
\section{Velocity induced by the vortex ring}
\label{s:velVortexRing}
%=========================================================================
The motion of the gas in cluster around the toroidal bubble is not
irrotational.  While the results of the previous section remain valid,
they are incomplete.  To describe motion of the gas around a toroidal
bubble we have to take into account the rotation of the gas around the
centre line of the torus.  The rotation itself does not change the
value of the added-mass, but it is essential for proper treatment of
the velocity field around the toroidal bubble.

The following general discussion is based on the material from
\cite{Batchelor67}.

Consider an incompressible fluid with the velocity field, $\bm{v}$,
\[
\nabla\cdot\bm{v}=0,
\]
which can be described in terms of the vector potential, $\bm{B}$,
\begin{equation}
\label{e:velvr1}
\bm{v} = \nabla\times\bm{B},
\end{equation}
and vorticity vector $\bomega$,
\[
\bomega = \nabla\times\bm{v}.
\]
The above equations imply
\begin{equation}
\label{e:velvr2}
\nabla\left(\nabla\cdot\bm{B}\right)-\nabla^2\bm{B}=\bomega.
\end{equation}
Se can seek a solution of (\ref{e:velvr2}) in the volume where
$\nabla\cdot\bm{B}=0$, so that,
\[
\nabla^2\bm{B}=-\bomega,
\]
and
\begin{equation}
\label{e:velvr3}
\bm{B}(\bm{x}) = \frac{1}{4\pi}\int\frac{\bomega(\bm{x}')}{s}\dd
V(\bm{x}'),
\end{equation}
where $s=|\bm{x}-\bm{x}'|$.  Using (\ref{e:velvr3}) the gauge
$\nabla\cdot\bm{B}=0$ reduces to $\bomega\cdot\bm{n}=0$, where
$\bm{n}$ is the normal vector to the boundary of the fluid.  The
latter can always be satisfied in a volume possibly extended beyond
the physical boundary \citep[][ \S 2.4]{Batchelor67}.  Using
(\ref{e:velvr1}) we find the velocity of the fluid in terms of the
vorticity vector,
\begin{equation}
\label{e:velvr4}
\bm{v}(\bm{x}) =
-\frac{1}{4\pi}\int\frac{\bm{s}\times\bomega(\bm{x}')}{s^3}\dd
V(\bm{x}').
\end{equation}

The {\it vortex-line} (by analogy with the flow line) is the line
whose tangent vector is everywhere parallel to the local value of the
vorticity $\bomega$.  The {\it vortex-tube} is a surface constructed
by all vortex lines passing through a given reducible closed curve
$C$\footnote{This definition of the vortex-tube is identical to the
definition of the flux tubes in a magnetic field.  Further analogies do
also hold, {\it e.g.}, (\ref{e:velvr6}) is analogues to the Biot--Savart
law, where the line vortex plays the role of a current, and velocity
plays the role of the resulting magnetic field.}.  The integral of
vorticity over an open surface $A$ bounded by the same closed curve
$C$, lying on a vortex tube and passing round it once is independent
of the choice of the line and, therefore, the surface $A$.  In other
words, the flux of vorticity through the vorticity-tube, along the
vortex-line, is conserved.  The value of the surface integral over the
surface $A$, i.e. the flux of vorticity through a cross section of the
vortex-tube, is called the strength of the vortex tube,
\[
\Gamma =\int_A \bomega\cdot\bm{n}\dd A=\oint_C\bm{v}\cdot\dd x.
\]
The line integral is taken over a closed line, $C$, lying on the
vortex-tube and passing around it once.  It is called {\it
circulation}.

The {\em line vortex} is a useful mathematical abstraction in cases when
the vorticity is large in a limited volume (in the vicinity of a line)
and negligible elsewhere.  If $\delta\bm{l}$ is a vector element of
the line vortex which lies in the volume $\delta V$, by the definition we
have,
\begin{equation}
\label{e:velvr5}
\int_{\delta V}\bomega\dd V = \Gamma \delta\bm{l},
\end{equation}
where $\Gamma$ is the strength of the vortex tube constructed around
the line vortex.  Using (\ref{e:velvr5}) and (\ref{e:velvr4}) it is
straightforward to get the velocity distribution of the fluid,
\begin{equation}
\label{e:velvr6}
\bm{v}(\bm{x}) =
-\frac{\Gamma}{4\pi}\oint\frac{\bm{s}\times\dd\bm{l}(\bm{x}')}{s^3},
\end{equation}
where $\bm{s}$ is the vector connecting the point $\bm{x}$ in the flow
with the point $\bm{x}'$ on the line vortex.

%------------------------------------------------------------------
\subsection{Application to buoyant bubbles}
\label{s:ATBB}
%------------------------------------------------------------------
We have argued (PI) that the velocity field of plasma induced by the
ascending bubbles is that of a vortex ring, \ie a circular line
vortex of radius $R$ and strength $\Gamma$.  Using (\ref{e:velvr6}) it
is easy to get an expression of the velocity of the plasma directly
under the rising circular line vortex (vortex ring),
\begin{equation}
\label{e:ATBB1}
v(x) = \frac{\Gamma R^2}{2(x^2+R^2)^{3/2}},
\end{equation}
where $x$ is the distance from the centre of the bubble to the point
directly under it.  This is the velocity with which the material will
be lifted up from the centre of the cluster by the ascending bubble
due to the circulation.

The velocity field induced by the vortex ring has another interesting
property.  It is obvious from (\ref{e:velvr6}) that there is a
singularity in the velocity field at the points on the vortex.
Careful examination of the velocity field in the vicinity of the line
vortex \citep{Batchelor67} shows that the velocity field near the
vortex consists of the rotational motion around it plus a
translational motion.  If $\bm{t}$, $\bm{n}$, $\bm{b}$ are the
tangent, the normal, and the binormal vectors of the line vortex at
the point $P$, then the position of a point in the plane
perpendicular to the line vortex at $P$ can be written as,
\[
\bm{r} = q_2\bm{n} + q_3\bm{b}.
\]
The velocity field at the distance $r=\sqrt{q_2^2+q_3^2}\rightarrow0$
near the point $P$, see Fig.~\ref{f:sch}, has the asymptotics
\citep[][ \S~7.1]{Batchelor67},
\begin{equation}
\label{e:ATBB2}
\frac{\Gamma}{2\pi r} \left(\bm{b}\cos\phi-\bm{n}\sin\phi\right)
+\frac{\Gamma}{4\pi R}\bm{b}\log\frac{R}{r}+O\left(r^0\right),
\end{equation}
where $\phi$ is a polar angle in the plane defined by the vectors
$\bm{n}$ and $\bm{b}$.  The first term in (\ref{e:ATBB2}) represents
the expected circulatory motion about the line vortex. The
second term shows that there is another weaker singularity of the
velocity distribution associated with the local curvature of the line
vortex, this is the induced velocity,
\[
\begin{split}
\bm{v}_\text{ind}&=\bm{b}v_\text{ind},\\
v_\text{ind} &= \frac{\Gamma}{4\pi
  R}\bm{b}\log\frac{R}{r}+O\left(r^0\right).
\end{split}
\]
The velocity of the fluid in the vicinity of point $P$ has a
large velocity in the direction of the binormal, with a magnitude
varying asymptotically as $\log 1/r$.

\cite{Lamb32} (Ch.~7) gives a proof of the Kelvin formula for the self
induced velocity of a circular vortex ring of a small cross section in
a perfect fluid,
\begin{equation}
\label{e:ATBB3}
v_\text{ind} = \frac{\Gamma}{4\pi R} \biggl[\log\frac{8R}{r} -
\frac{1}{4} + o\left(\frac{r}{R}\right)\biggr].
\end{equation}
In case when the vorticity is confined to the surface of the ring (so
called hollow vortex), the term $-1/4$ is replaced by $-1/2$ \citep{Hicks1883}.

Equation (\ref{e:ATBB3}) was applied to buoyant vortex rings by
\cite{Turner57} \citep[see also][]{Morton60,Turner69,Woods97}.  The
general problem was re-examined by \cite{Fraenkel72}, who proved the
existence of steady vortex rings, and gave a generalisation of
(\ref{e:ATBB3}) for arbitrary distributions of vorticity as a function
of $r$, and \cite{Saffman70} for the case of viscous vortex rings.
\cite{Onufriev67} used the following formula for the induced velocity
for description of motion of the cloud from a nuclear explosion,
\begin{equation}
\label{e:ATBB4}
\begin{split}
v_\text{ind}\approx\frac{\Gamma}{4\pi R} \biggl[\log\frac{8R}{r} -
\frac{1}{4}\\
-\frac{12\log(8R/r)-15}{32}\left(\frac{r}{R}\right)^2\biggr].
\end{split}
\end{equation}

\bibliography{gbp}

\end{document}